\documentclass[aps,prl,twocolumn,showpacs,groupedaddress]{revtex4}
\usepackage{graphicx}
\usepackage{amssymb}
\newlength{\figwidth}

\usepackage{epsf}
\begin{document}
\preprint{\today}
\title{Scanning Gate Microscopy of a Nanostructure where Electrons Interact}

\author{Axel Freyn}
%
\affiliation{Service de Physique de l'\'Etat Condens\'e (CNRS URA 2464), 
DSM/IRAMIS/SPEC, CEA Saclay, 91191 Gif-sur-Yvette Cedex, France}
\author{Ioannis Kleftogiannis}
\altaffiliation[Present address: ]{Department of 
Physics, University of Ioannina, Ioannina 45 110, Greece}

\affiliation{Service de Physique de l'\'Etat Condens\'e (CNRS URA 2464), 
DSM/IRAMIS/SPEC, CEA Saclay, 91191 Gif-sur-Yvette Cedex, France}

\author{Jean-Louis Pichard}
%
\affiliation{Service de Physique de l'\'Etat Condens\'e (CNRS URA 2464), 
DSM/IRAMIS/SPEC, CEA Saclay, 91191 Gif-sur-Yvette Cedex, France}

\begin{abstract}  
We show that scanning gate microscopy can be used for probing 
electron-electron interactions inside a nanostructure. We assume 
a simple model made of two non interacting strips attached 
to an interacting nanosystem. In one of the strips, the 
electrostatic potential can be locally varied by a charged tip.  
This change induces corrections upon the nanosystem Hartree-Fock 
self-energies which enhance the fringes spaced by half the Fermi 
wave length in the images giving the quantum conductance as a 
function of the tip position. 
\end{abstract}


\maketitle

\begin{figure}
\centerline{
\epsfxsize=\columnwidth
\epsfysize=2.5cm 
\epsffile{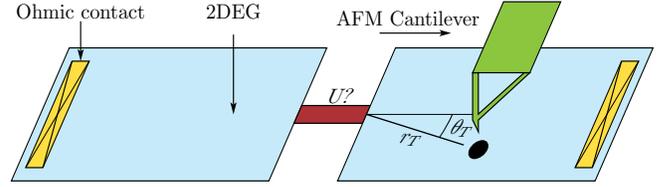}
}
\caption{Scheme of a SGM setup: Two 2DEGs are connected via a 
nanostructure (red). The negatively (positively) charged tip 
creates a small depletion (accumulation) region ($\bullet$) which 
scatters electrons in the right 2DEG. By scanning the tip and measuring 
the quantum conductance $g$ between the 2 ohmic contacts, one can detect 
the interaction $U$ acting inside the nanostructure.}  
\label{fig1} 
\end{figure}

Semiconductor nanostructures based on two dimensional electron gases (2DEGs) 
have been extensively studied, with the expectation of developing future 
devices for sensing, information processing and quantum computation. 
Scanning gate microscopy (SGM) consists in using the charged tip of an AFM 
cantilever as a movable gate for studying these nanostructures. A typical SGM 
setup is sketched in FIG.~\ref{fig1}. A negatively charged tip capacitively 
couples with the 2DEG at a distance $r_T$ from the nanostructure, creating 
a small depletion region that scatters the electrons. Scanning the tip 
around the nanostructure and measuring the quantum conductance $g$ between 
two ohmic contacts put on each side of the nanostructure as a function of 
the tip position provide the SGM images. If the nanostructure is a quantum 
point contact (QPC), the charged tip can reduce \cite{topinka2001} $g$ 
by a significant fraction $\delta g=g-g_0$, when the conductance without 
tip $g_0$ is biased on the first conductance plateau $g_q=2e^2/h$. 
Moreover, fringes spaced by $\lambda_F/2$, half the Fermi wave length, and 
falling  off with distance $r_T$ from the QPC, can be seen in the 
experimental images giving $\delta g$ as a function of the tip position. 
Very small distances $r_T$ were not scanned in Refs.~\cite{topinka2000,
topinka2001}, but this was done \cite{aoki} later, giving extra ring 
structures inside the QPC if $g_0$ is biased between the conductance 
plateaus. Scanning gate microscopy has been recently used for studying 
QPCs \cite{jura},  open quantum rings \cite{sellier} and quantum dots 
created in carbon nanotubes \cite{mcEuen} and 2DEGs \cite{ensslin}. 

Many features of the observed SGM images can be described by single 
particle theories \cite{heller,bruno,sellier}. However, many body 
effects are expected to be important inside certain nanostructures 
(almost closed QPC around the $0.7 (2e^2/h)$ conductance anomaly 
\cite{pepper}, quantum dots of low electron density). We show in this 
letter that these many body effects can be observed in the SGM images 
of such nanostructures. Two main signatures of the interaction are 
identified: fringes of enhanced magnitude, falling 
off as $1/r_T^2$ near the nanostructure, before falling off as $1/r_T$ 
far from the nanostructure, and a phase shift of the fringes between 
these two regions. Though we study this interaction effect using a 
very simple model, our theory can be extended to any nanostructure 
inside which electrons interact. 

Without interaction, the nanostructure and the depletion region created by 
the tip are independent scatterers. With interactions inside the 
nanostructure, the effective nanostructure transmission becomes non 
local and can be modified by the tip. The origin of this non local effect 
is easy to explain \cite{afp,fp1,fp2} if one uses the Hartree-Fock (HF) 
approximation. The tip induces Friedel oscillations of the electron density, 
which can modify the density inside the nanostructure. As one moves the tip, 
this changes the Hartree corrections of the nanostructure. A similar effect 
changes also the Fock corrections \cite{afp,fp1,fp2}. When the electrons do 
not interact inside the nanostructure, the SGM images probe the 
interferences of electrons which are transmitted by the nanostructure 
and elastically backscattered by the tip at the Fermi energy $E_F$. When 
the electrons interact inside the nanostructure, the information given by 
the SGM images becomes more complex, since the scattering processes of 
energies below $E_F$ influence also the quantum conductance. In the HF 
approximation, these non local processes taking place at all energies 
below $E_F$ are taken into account by the integral equations giving the 
nanosystem HF corrections. 

The principle for the detection of the interaction $U$ via SGM can be simply 
explained in one dimension, when the strips are semi-infinite 
chains. If $U=0$, the transmitted flow interferes with the flow reflected 
by the tip, giving rise to Fabry-P\'erot oscillations which do not decay 
as $r_T \rightarrow \infty$. Hence the conductance $g$ of a 
nanostructure in series with a tip exhibits oscillations which do not 
decay when $r_T$ increases. If $U \neq 0$, the HF-corrections of the 
nanostructure are modified by the Friedel oscillations induced by the 
tip inside the nanostructure. This gives an additional effect for $g$, 
which decays as the Friedel oscillations causing it ($1/r_T$-decay in 1d, 
with oscillations of period $\lambda_F/2$). Measuring $g$ as a function 
of the tip position, one gets oscillations of period $\lambda_F/2$ in 
the two cases, but their decays are different and allow to measure the 
interaction strength $U$ inside the nanosystem.    

 Interactions in 1d chains give rise to a Luttinger-Tomonaga liquid and 
cannot be neglected. It is necessary to take 2d strips of sufficient 
electron density (small factor $r_s$) for neglecting interaction outside 
the nanosystem. The effect of the tip becomes more subtle with 2d strips:  
First, the Friedel oscillations decreasing as $1/r^d$ in $d$ dimensions, 
the effect of the tip upon $g$ has a faster decay, unless focusing effects 
take place. Second, the non interacting limit becomes more complicated. 
The probability for an electron of energy $E_F$ to reach the tip, and to 
be reflected through the nanostructure also decays as $r_T \rightarrow 
\infty$. Assuming isotropy, the probabilities of these two events should 
decay as $1/r_T$, giving a total $1/r^2_T$ decay for $g$. But isotropy 
is not a realistic assumption for SGM setups. The transmission can be 
strongly focused, making the effect of the tip a function of the angle 
$\theta_T$. Spectacular focusing effects have been observed 
\cite{topinka2000} using a QPC: The effect of the tip is mainly 
focused around $\theta_T \approx 0$  or $\pm \pi/4$, depending if 
$g_0\approx g_q$ or $2 g_q$.

\begin{figure}
\centerline{
\epsfxsize=\columnwidth 
\epsfysize=5cm 
\epsffile{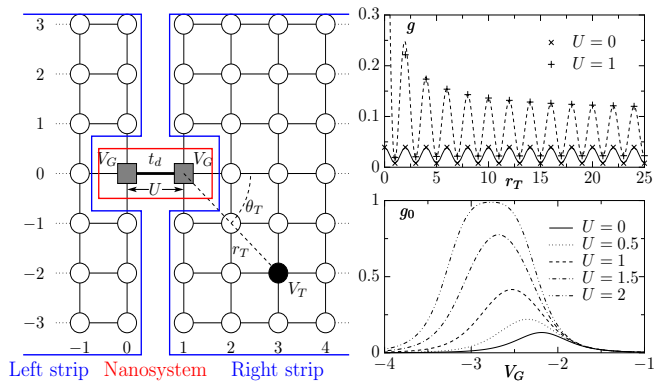}
}
\caption
{
Left: Used model: 2 strips of width $2 L_y+1$ (here $L_y=3$) are 
connected via a nanosystem (2 sites $\blacksquare$,  hopping $t_d$ and potentials 
$V_G$). The repulsion $U$ acts only inside the nanosystem. The charged tip 
gives rise to a potential $V_T$ ($\bullet$) at a distance $r_T$ from the 
nanosystem. Upper right: SGM measure using 1d chains ($L_y=0$) at 
half-filling ($k_F=\pi/2$). The conductance $g$ of the nanosystem 
($V_G=-U/2$ and $t_d=0.1$) in series with a tip ($V_T=2$) is given 
as a function of $r_T$. Fits $0.024 + 0.016 \cos (\pi r_T)$ (solid line) 
and $0.066+0.132/r_T+(0.043 + 0.014/r_T)\cos (\pi r_T)$ (dashed line). 
Lower right: Conductance $g_0$ without tip ($V_T=0$) as a function of 
$V_G$ for 2d strips ($2 L_y+1=301$). $E_F=-3.57$ ($k_F=0.668$), $t_d=0.1$ 
and different values of $U$. 
}
\label{fig2} 
\end{figure}

For studying SGM with 2d strips more precisely, we use a simple model 
sketched in FIG.~\ref{fig2} (left), assuming spin polarized electrons 
(spinless fermions). The Hamiltonian reads $H=H_{\text{nano}}+ 
H_{\text{strips}}+H_{T}$. For the nanostructure, we take a nanosystem with 
two sites of energy $V_G$ and of hopping term $t_d$. For the interaction, 
we take a repulsion of strength $U$ between these two sites. We assume that 
$V_G$ can be varied by an external gate. The Hamiltonian of the nanosystem 
reads
\begin{equation}
H_{\text{nano}}= V_G \sum_{x=0}^1 n_{x,0} - t_d 
(c^\dagger_{0,0} c^{\phantom{\dagger}}_{1,0} + H.c.) + 
U n_{0,0} n_{1,0}. 
\end{equation}
$c^{\phantom{\dagger}}_{x,y}$ ($c^\dagger_{x,y}$) is the annihilation 
(creation) operator at site $x,y$, and $n_{x,y} = c^\dagger_{x,y} 
c^{\phantom{\dagger}}_{x,y}$.
\begin{equation}
H_{\text{strips}} = -t_{\mathrm h} \sum_{x,y} (c^{\dagger}_{x,y} c_{x,y+1} + 
c^{\dagger}_{x,y} c_{x+1,y} + H.c.) 
\end{equation}
describes the strips and their couplings to the nanosystem (see 
FIG.~\ref{fig2} (left)). We assume hard wall boundaries in the $y$-direction. 
$t_{\mathrm h}=1$ sets the energy scale. The depletion (accumulation) 
region created by a negatively (positively) charged tip located on top of a 
site of coordinates $(x_T>1,y_T)=r_T (\cos \theta_T, \sin\theta_T)$ is 
described by a local Hamiltonian $H_{T}=V_T n_{x_T,y_T}$.

 In FIG.~\ref{fig2} (upper right), we show how to detect $U$ by scanning 
gate microscopy in the 1d limit of our model ($L_y=0$). The chains are 
half-filled ($E_F=0$), and the conductance $g$ of the nanosystem in series 
with a tip is given as a function of $r_T$. If $U=0$, $g$ exhibits even-odd 
oscillations of constant amplitude, while these oscillations fall off as 
$1/r_T$ near the nanosystem if $U \neq 0$. When $L_y=0$, the HF corrections 
can be obtained using an extrapolation method \cite{afp,fp1,fp2}. When $L_y$ 
is large, using self-energies becomes more efficient for calculating the HF 
corrections and the conductance $g$. 

 The retarded ($z=E+i\eta$) and advanced ($z=E-i\eta$) Green's functions 
of the nanosystem at an energy $E$, $\eta \to 0^+$ are given by the 
$2 \times 2$-matrix 
\begin{equation} 
\label{gnano}
G_{\text{nano}}= \left (
\begin{array}{cc}
z-V_G-\sigma_{\mathbf 0}-\Sigma^{\text{H}}_{\mathbf 0} & 
t_d - \Sigma^{\text{F}} \\
t_d - \Sigma^{\text{F}} &
z-V_G-\sigma_{\mathbf 1}-\Sigma^{\text{H}}_{\mathbf 1}
\end{array} \right)^{-1}.
\end{equation}
The self-energies $\sigma_{\mathbf 0}$ and $\sigma_{\mathbf 1}$ describe  
the couplings of the left and right strips to the nanosystem sites 
$\mathbf 0=(0,0)$ and $\mathbf 1=(1,0)$ respectively. If 
$G_{\text{strip}}^{L,R}$ are the Green's functions of the two strips 
excluding the 2 nanosystem sites, one gets 
\begin{eqnarray}
\sigma_{\mathbf 0}=& \sum_{I,J} \langle I|G_{\text{strip}}^{L} |J \rangle \\
\sigma_{\mathbf 1}=& \sum_{I,J} \langle I|G_{\text{strip}}^{R} |J \rangle.
\label{sigma}
\end{eqnarray}
$I$ and $J$ label the sites $(-1,0),(0,1)$ and $(0,-1)$ directly coupled 
to $\mathbf 0$ for $\sigma_{\mathbf 0}$, and the sites $(2,0),(1,1)$ and 
$(1,-1)$ directly coupled to $\mathbf 1$ for $\sigma_{\mathbf 1}$. For 
each tip position and different energies $E \leq E_F$, the Green's 
functions of the right strip determining  $\sigma_{\mathbf 1}$ are 
calculated using recursive Green's function (RGF) algorithm 
(see Ref.~\cite{bruno} and references therein). 

The self-energies $\Sigma^{\text{H}}_{\mathbf 0}$ and 
$\Sigma^{\text{H}}_{\mathbf 1}$ describe the Hartree corrections 
yielded by the inter-site repulsion $U$ to the potentials of the sites 
$\mathbf 0$ and $\mathbf 1$ respectively, while the Fock self-energy 
$\Sigma^{\text{F}}$ 
modifies the hopping term $t_d$ because of exchange. The matrix elements 
$(G_{\text{nano}}(E))_{i,j}$ ($i,j=0,1$) being given by Eq.~(\ref{gnano}), 
the HF self-energies are the self-consistent solution of 
3 coupled integral equations:
\begin{eqnarray}
\Sigma^{\text{H}}_{\mathbf 0} &= -\frac{U}{\pi} 
\Im \int_{-\infty}^{E_F} (G_{\text{nano}}(E))_{1,1} dE 
\label{HF-equation1}
\\
\Sigma^{\text{H}}_{\mathbf 1} &=-\frac{U}{\pi} 
\Im \int_{-\infty}^{E_F} (G_{\text{nano}}(E))_{0,0} dE 
\label{HF-equation2}
\\
\Sigma^{\text{F}} &=\frac{U}{\pi} 
\Im \int_{-\infty}^{E_F} (G_{\text{nano}}(E))_{0,1} dE.
\label{HF-equation3}
\end{eqnarray}

 The imaginary parts of the above integrals are equal to zero for $E<-4$. 
For $-4<E<E_F$, the poles on the real axis make necessary to integrate 
Eqs.~(\ref{HF-equation1}-\ref{HF-equation3}) using Cauchy theorem. We have 
used a semi-circle centered at $(E_F-4)/2$ in the upper part of the complex 
plane. The integration is done using the Gauss--Kronrod algorithm. 
This requires to calculate $G_{\text{nano}}$ (and therefore 
$\sigma_{\mathbf 0}(z)$ and $\sigma_{\mathbf 1}(z,V_T)$) for a sufficient 
number  ($\approx 100$) of complex energies $z$ on the semi-circle, before 
determining the self-consistent solutions of Eqs.~(\ref{HF-equation1}-
\ref{HF-equation3}) recursively.  Calculating $\sigma_{\mathbf 1}(z,V_T)$ 
for each tip position ($r_T,\theta_T$), one can obtain the 2d images 
giving $\Sigma^{\text{HF}}$ as a function of the tip position.

Once the self-energies $\Sigma^{\text{HF}}$ are obtained in the zero 
temperature limit, the interacting nanosystem is described by an effective 
one body Green's function, identical to the one of a non interacting 
nanosystem, with potentials $V_G+\Sigma^{\text{H}}_{\mathbf 0}$ and $V_G+
\Sigma^{\text{H}}_{\mathbf 1}$ and hopping $-t_d+\Sigma^{\text{F}}$.  
Then, the many channel Landauer-Buttiker formula $g=\text{trace } 
tt^{\dagger}$ valid for non interacting systems can be used to obtain 
the zero temperature conductance $g$ in units of $e^2/h$ (for polarized 
electrons). This conductance corresponds to a measure made between the 
two ohmic contacts sketched in FIG.~\ref{fig1}. We use the RGF algorithm 
to obtain the  Green's function of the measured system, from which the 
transmission matrix $t$ can be expressed \cite{datta}. 

 For having negligible lattice effects and SGM images characteristic of 
 the continuum limit, we consider a low filling factor $\nu \approx 1/25$
in the 2d strips, corresponding to a Fermi energy (momentum) $E_F=-3.57$ 
($k_F=0.668$). The width of the strip ($2L_y+1=301$) is sufficient for 
having a 2d behavior in the vicinity of the nanosystem. Moreover, we take 
small values of the nanosystem hopping $t_d$, in order to increase 
\cite{fp1,fp2} the effect of the tip upon the HF self-energies. In 
FIG.~\ref{fig2} (lower right), the conductance $g_0$ without tip ($V_T=0$) 
is given as a function of the gate potential $V_G$ for increasing values 
of $U$. When $t_d$ is small, the double peak structure of $g_0(V_G)$ 
characteristic of a nanosystem with two sites merges \cite{fp2} to form a 
single peak. Hereafter, the SGM images are given for a gate potential 
$V_G^*(U)$ for which $g_0(V_G)$ is maximum.  

\begin{figure}
\centerline{
\epsfxsize=\columnwidth
\epsffile{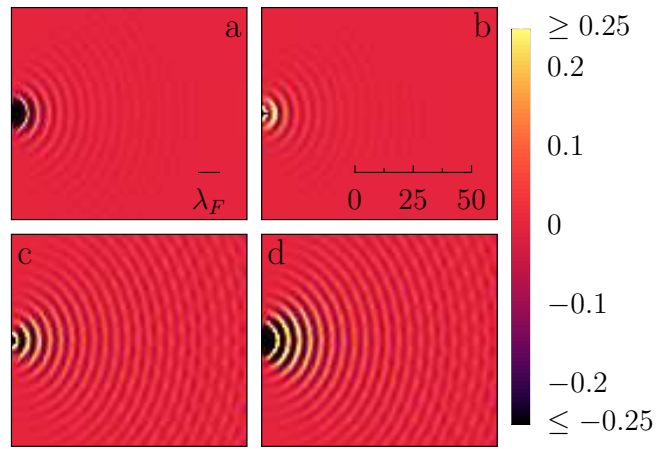}
}
\caption{Images obtained by scanning the tip ($V_T=-2$) on the 
right strip. The Fermi wave length $\lambda_F$ and the scale 
are given in FIG.~(a) and (b). Upper part: Relative corrections 
$\delta \Sigma^{\text{F}}/\Sigma^{\text{F}}(V_T=0)$ (FIG.~a) and 
$\delta \Sigma^{\text{H}}_{\mathbf 0}/\Sigma^{\text{H}}_{\mathbf 0}(V_T=0)$ 
(FIG.~b) of the Fock and Hartree self-energies as a function of the 
tip position  for $U=1.7$. $\delta \Sigma=\Sigma(V_T=-2)-\Sigma (V_T=0)$. 
$\Sigma^{\text{F}} (V_T=0)=-0.120$ and $\Sigma^{\text{H}}_{\mathbf 0}
(V_T=0)=0.529$.Lower part: Relative corrections $\delta g/g_0$ as a 
function of the tip position for $U=0$ (FIG.~c) and $U=1.7$ (FIG.~d). 
Used parameters: $E_F=-3.57$, $t_d=0.01$ and strips of width $2L_y+1=301$. 
$V_G=V_G^*=-2.870$  ($-2.187$) and $g_0=0.188$ ($0.0014$) for $U=1.7$ 
($U=0$).
}
\label{fig3} 
\end{figure} 
\begin{figure}
\centerline{
\epsfxsize=\columnwidth
\epsfysize=7cm 
\epsffile{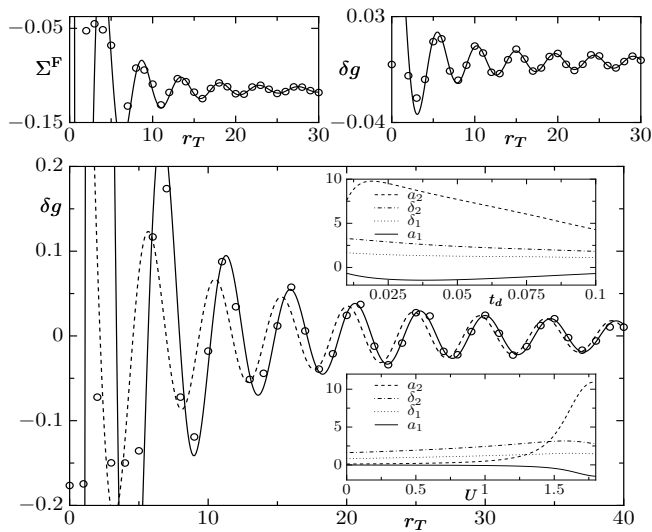}
}
\caption{Effect of the tip as a function of $r_T$ for $\theta_T=0$ 
when $E_F=-3.57$, $2L_y+1=301$ and $V_T=-2$. Upper left: $1/r_T^2$ 
decay of the Fock self-energy $\Sigma^{\text{F}}$ ($\circ$) for 
$U=1.7$.  Solid line: $-0.115+2.379 \cos (2k_Fr_T+ 0.765)/r_T^2$. 
Same parameters as in FIG.~\ref{fig3} (a). Upper right: $1/r_T$-decay of 
$\delta g$ for $U=0$ and $t_d=0.1$, $g_0=0.133$, $V_G^*=-2.187 $. 
Solid line: $0.110 \cos (2 k_F r_T -1.2)/r_T$. Lower part: 
$1/r_T^2$-decay of $\delta g$ ($\circ$) at small distances $r_T$ 
induced by an interaction $U=1.7$ when $t_d=0.01$ ($g_0=0.188$). 
Fit $F(r_T)$ with $a_1=-0.676$, $a_2=-7.605$, $\delta_1=1.664$ and 
$\delta_2=0.120$ (solid line). Taking $a_2=0$ in $F(r_T)$ (dashed line) 
fails to describe $\delta g$ at small distances $r_T$, and does not 
matter at large distances $r_T$. Insets: Parameters of $F(r_T)$ fitting 
$\delta g$ as a function of $t_d$ when $U=1.7$ (upper inset) and as a 
function of $U$ when $t_d=0.02$ 
(lower inset). 
}
\label{fig4} 
\end{figure}

 The effect of the tip upon $\Sigma^{\text{F}}$ and 
$\Sigma^{\text{H}}_{\mathbf 0}$ is shown  as a function of the tip position 
($x_T,y_T$) in the upper part of FIG.~\ref{fig3}. The images show fringes 
spaced by $\lambda_F/2$ which fall off as $1/r_T^2$. In FIG.~\ref{fig4} 
(upper left), the Fock term 
$\Sigma^{\text{F}}$ is plotted as a function of $r_T$ for $\theta_T=0$. 
The decay can be described by a $\cos(2k_Fr_T+\delta)/r_T^2$ fit. Similar 
fits characterize the 2 Hartree terms. Since the effect of the tip upon 
$\Sigma^{\text{H}}_{\mathbf 0}$ is driven by Friedel oscillations, 
$\Sigma^{\text{H}}_{\mathbf 0}$ decays as 2d Friedel oscillations. 
$\Sigma^{\text{H}}_{\mathbf 1}$ and $\Sigma^{\text{F}}$ have similar decays.

  In the lower part of FIG.~\ref{fig3}, the effect of the tip upon 
the conductance is given as a function of ($x_T,y_T$). 
The left figure gives $\delta g/g_0$  without interaction ($U=0$), 
where $\delta g=g(V_T=-2)-g_0$. One can see that $\delta g$ decays as 
$r_T$ increases, the image exhibiting fringes spaced by $\lambda_F/2$. 
The decay depends on the angle $\theta_T$. For $\theta_T=0$, 
$\delta g(U=0)$ falls off as $1/r_T$, and not as $1/r_T^2$ 
(isotropic assumption). This is shown in FIG.~\ref{fig4} (upper right), 
a fit of the form $a_1 \cos(2k_Fr_T+\delta_1)/r_T$ describing the decay.
 
 In FIG.~\ref{fig3} (d), $\delta g/g_0$ is shown when the electrons 
interact inside the nanosystem ($U=1.7$). The interation effect 
$\propto 1/r_T^2$ of the tip upon $g$ via $\Sigma^{\text{HF}}(V_T)$ 
enhances the fringes near the nanosystem. Since the SGM images exhibit 
fringes spaced by $\lambda_F/2$, decaying as $1/r_T^2$ for 
$\Sigma^{\text{HF}}$ and as $1/r_T$ for $g$ when $U=0$, we fit the 
effect of the tip upon $g$ when $U \neq 0$ by a function 
\begin{equation}
F(r_T)=\frac{a_1 \cos(2k_Fr_T+\delta_1)}{r_T}
+\frac{a_2 \cos(2k_Fr_T+\delta_2)}{r_T^2}  
\label{fit}
\end{equation} 
which contains 4 adjustable parameters $a_1,\delta_1,a_2$ and $\delta_2$. 
The $\cos(2k_Fr_T+\delta)$ terms give the fringes. The $1/r_T$ term fits 
the decay without interaction. The $1/r_T^2$ term is added for taking 
into account the effect of the tip upon $g$ occurring near the nanosystem 
via $\Sigma^{\text{HF}}$. 
 
 When $r_T$ is large, the effect of the tip upon $\Sigma^{\text{HF}}$ 
is negligible and $\delta g(r_T)$ falls off as when $U=0$ ($1/r_T$ decay). 
As $r_T$ varies, a crossover from a decay described by the term 
$a_2 \cos(2k_Fr_T+\delta_2)/r_T^2$ of $F(r_T)$ towards a decay described 
by $a_1 \cos(2k_Fr_T+\delta_1)/r_T$ takes place. This is shown in 
FIG.~\ref{fig4} (lower part), where $a_2\neq 0$ is necessary for 
describing $g(r_T)$ at small distances $r_T$. This crossover is accompanied 
by a phase shift of the fringes ($\delta_2 \neq \delta_1$), which can 
be seen in the figure. To find this crossover, we have studied how the 
4 parameters of $F(r_T)$ depend on $t_d$ and on $U$  
(insets of FIG.~\ref{fig4}). To get a $1/r_T^2$ decay which 
persists in a large domain around the nanosystem, one needs $|a_2| \gg 
|a_1|$. This occurs for small $t_d$ (upper inset) and $U>1$ 
(lower inset).

  In summary, neglecting electron-electron interactions and disorder in the 
strips, we have shown that the SGM images allow to measure the 
interaction strength inside the nanosystem. From 
zero temperature transport measurements, one can detect a $1/r_T^2$ 
decay of the SGM images around the nanosystem, and via $a_2(U)$, the 
value of $U$ characteristic of the nanosystem can be determined. For 
observing this $1/r_T^2$ decay, one needs (i) large electron-electron 
interactions inside the nanostructure (sufficient $r_s$ factor), (ii) 
large density oscillations induced by the tip ($V_T$ large), and (iii) 
that those oscillations modify the density inside the nanostructure 
($r_T$ not too large, strong coupling between the nanostructure and the 
strips). 

%
%
\begin{acknowledgments}
We thank S.~N.~Evangelou for useful comments and a careful reading of the 
manuscript. The support of the network ``Fundamentals of nanoelectronics'' 
of the EU (contract MCRTN-CT-2003-504574) is gratefully acknowledged.
\end{acknowledgments}


\begin{thebibliography}{14}
\expandafter\ifx\csname natexlab\endcsname\relax\def\natexlab#1{#1}\fi
\expandafter\ifx\csname bibnamefont\endcsname\relax
  \def\bibnamefont#1{#1}\fi
\expandafter\ifx\csname bibfnamefont\endcsname\relax
  \def\bibfnamefont#1{#1}\fi
\expandafter\ifx\csname citenamefont\endcsname\relax
  \def\citenamefont#1{#1}\fi
\expandafter\ifx\csname url\endcsname\relax
  \def\url#1{\texttt{#1}}\fi
\expandafter\ifx\csname urlprefix\endcsname\relax\def\urlprefix{URL }\fi
\providecommand{\bibinfo}[2]{#2}
\providecommand{\eprint}[2][]{\url{#2}}

\bibitem[{\citenamefont{Topinka et~al.}(2001)\citenamefont{Topinka, LeRoy,
  Westervelt, Shaw, Fleischmann, Heller, Maranowski, and
  Gossard}}]{topinka2001}
\bibinfo{author}{\bibfnamefont{M.~A.} \bibnamefont{Topinka}},
  \bibinfo{author}{\bibfnamefont{B.~J.} \bibnamefont{LeRoy}},
  \bibinfo{author}{\bibfnamefont{R.~M.} \bibnamefont{Westervelt}},
  \bibinfo{author}{\bibfnamefont{S.~E.~J.} \bibnamefont{Shaw}},
  \bibinfo{author}{\bibfnamefont{R.}~\bibnamefont{Fleischmann}},
  \bibinfo{author}{\bibfnamefont{E.~J.} \bibnamefont{Heller}},
  \bibinfo{author}{\bibfnamefont{K.~D.} \bibnamefont{Maranowski}},
  \bibnamefont{and} \bibinfo{author}{\bibfnamefont{A.~C.}
  \bibnamefont{Gossard}}, \bibinfo{journal}{Nature}
  \textbf{\bibinfo{volume}{410}}, \bibinfo{pages}{183} (\bibinfo{year}{2001}).

\bibitem[{\citenamefont{Topinka et~al.}(2000)\citenamefont{Topinka, LeRoy,
  Shaw, Heller, Westervelt, Maranowski, and Gossard}}]{topinka2000}
\bibinfo{author}{\bibfnamefont{M.~A.} \bibnamefont{Topinka}},
  \bibinfo{author}{\bibfnamefont{B.~J.} \bibnamefont{LeRoy}},
  \bibinfo{author}{\bibfnamefont{S.~E.~J.} \bibnamefont{Shaw}},
  \bibinfo{author}{\bibfnamefont{E.~J.} \bibnamefont{Heller}},
  \bibinfo{author}{\bibfnamefont{R.~M.} \bibnamefont{Westervelt}},
  \bibinfo{author}{\bibfnamefont{K.~D.} \bibnamefont{Maranowski}},
  \bibnamefont{and} \bibinfo{author}{\bibfnamefont{A.~C.}
  \bibnamefont{Gossard}}, \bibinfo{journal}{Science}
  \textbf{\bibinfo{volume}{289}}, \bibinfo{pages}{2323} (\bibinfo{year}{2000}).

\bibitem[{\citenamefont{Aoki et~al.}(2006)\citenamefont{Aoki, Burke, da~Cunha,
  Akis, Ferry, and Ochiai}}]{aoki}
\bibinfo{author}{\bibfnamefont{N.}~\bibnamefont{Aoki}},
  \bibinfo{author}{\bibfnamefont{A.}~\bibnamefont{Burke}},
  \bibinfo{author}{\bibfnamefont{C.~R.} \bibnamefont{da~Cunha}},
  \bibinfo{author}{\bibfnamefont{R.}~\bibnamefont{Akis}},
  \bibinfo{author}{\bibfnamefont{D.~K.} \bibnamefont{Ferry}}, \bibnamefont{and}
  \bibinfo{author}{\bibfnamefont{Y.}~\bibnamefont{Ochiai}},
  \bibinfo{journal}{J. Phys.: Conf. Series} \textbf{\bibinfo{volume}{38}},
  \bibinfo{pages}{79} (\bibinfo{year}{2006}).

\bibitem[{\citenamefont{Jura et~al.}(2007)\citenamefont{Jura, Topinka, Urban,
  Yazdani, Shtrikman, Pfeiffer, West, and Goldhaber-Gordon}}]{jura}
\bibinfo{author}{\bibfnamefont{M.~P.} \bibnamefont{Jura}},
  \bibinfo{author}{\bibfnamefont{M.~A.} \bibnamefont{Topinka}},
  \bibinfo{author}{\bibfnamefont{L.}~\bibnamefont{Urban}},
  \bibinfo{author}{\bibfnamefont{A.}~\bibnamefont{Yazdani}},
  \bibinfo{author}{\bibfnamefont{H.}~\bibnamefont{Shtrikman}},
  \bibinfo{author}{\bibfnamefont{L.~N.} \bibnamefont{Pfeiffer}},
  \bibinfo{author}{\bibfnamefont{K.~W.} \bibnamefont{West}}, \bibnamefont{and}
  \bibinfo{author}{\bibfnamefont{D.}~\bibnamefont{Goldhaber-Gordon}},
  \bibinfo{journal}{Nature} \textbf{\bibinfo{volume}{3}}, \bibinfo{pages}{841}
  (\bibinfo{year}{2007}).

\bibitem[{\citenamefont{Martins et~al.}(2007)\citenamefont{Martins, Hackens,
  Pala, Ouisse, Sellier, Wallart, Bollaert, Cappy, Chevrier, Bayot
  et~al.}}]{sellier}
\bibinfo{author}{\bibfnamefont{F.}~\bibnamefont{Martins}},
  \bibinfo{author}{\bibfnamefont{B.}~\bibnamefont{Hackens}},
  \bibinfo{author}{\bibfnamefont{M.~G.} \bibnamefont{Pala}},
  \bibinfo{author}{\bibfnamefont{T.}~\bibnamefont{Ouisse}},
  \bibinfo{author}{\bibfnamefont{H.}~\bibnamefont{Sellier}},
  \bibinfo{author}{\bibfnamefont{X.}~\bibnamefont{Wallart}},
  \bibinfo{author}{\bibfnamefont{S.}~\bibnamefont{Bollaert}},
  \bibinfo{author}{\bibfnamefont{A.}~\bibnamefont{Cappy}},
  \bibinfo{author}{\bibfnamefont{J.}~\bibnamefont{Chevrier}},
  \bibinfo{author}{\bibfnamefont{V.}~\bibnamefont{Bayot}},
  \bibnamefont{et~al.}, \bibinfo{journal}{Phys. Rev. Lett.}
  \textbf{\bibinfo{volume}{99}}, \bibinfo{pages}{136807}
  (\bibinfo{year}{2007}).

\bibitem[{\citenamefont{Woodside and McEuen}(2002)}]{mcEuen}
\bibinfo{author}{\bibfnamefont{M.~T.} \bibnamefont{Woodside}} \bibnamefont{and}
  \bibinfo{author}{\bibfnamefont{P.~L.} \bibnamefont{McEuen}},
  \bibinfo{journal}{Science} \textbf{\bibinfo{volume}{296}},
  \bibinfo{pages}{1098} (\bibinfo{year}{2002}).

\bibitem[{\citenamefont{Pioda et~al.}(2004)\citenamefont{Pioda, Kicin, Ihn,
  Sigrist, Fuhrer, Ensslin, Weichselbaum, Ulloa, Reinwald, and
  Wegscheider}}]{ensslin}
\bibinfo{author}{\bibfnamefont{A.}~\bibnamefont{Pioda}},
  \bibinfo{author}{\bibfnamefont{S.}~\bibnamefont{Kicin}},
  \bibinfo{author}{\bibfnamefont{T.}~\bibnamefont{Ihn}},
  \bibinfo{author}{\bibfnamefont{M.}~\bibnamefont{Sigrist}},
  \bibinfo{author}{\bibfnamefont{A.}~\bibnamefont{Fuhrer}},
  \bibinfo{author}{\bibfnamefont{K.}~\bibnamefont{Ensslin}},
  \bibinfo{author}{\bibfnamefont{A.}~\bibnamefont{Weichselbaum}},
  \bibinfo{author}{\bibfnamefont{S.~E.} \bibnamefont{Ulloa}},
  \bibinfo{author}{\bibfnamefont{M.}~\bibnamefont{Reinwald}}, \bibnamefont{and}
  \bibinfo{author}{\bibfnamefont{W.}~\bibnamefont{Wegscheider}},
  \bibinfo{journal}{Phys. Rev. Lett.} \textbf{\bibinfo{volume}{93}},
  \bibinfo{pages}{216801} (\bibinfo{year}{2004}).

\bibitem[{\citenamefont{Heller et~al.}(2005)\citenamefont{Heller, Aidala,
  LeRoy, Bleszynski, Kalben, Westervelt, Maranowski, and Gossard}}]{heller}
\bibinfo{author}{\bibfnamefont{E.~J.} \bibnamefont{Heller}},
  \bibinfo{author}{\bibfnamefont{K.~E.} \bibnamefont{Aidala}},
  \bibinfo{author}{\bibfnamefont{B.~J.} \bibnamefont{LeRoy}},
  \bibinfo{author}{\bibfnamefont{A.~C.} \bibnamefont{Bleszynski}},
  \bibinfo{author}{\bibfnamefont{A.}~\bibnamefont{Kalben}},
  \bibinfo{author}{\bibfnamefont{R.~M.} \bibnamefont{Westervelt}},
  \bibinfo{author}{\bibfnamefont{K.~D.} \bibnamefont{Maranowski}},
  \bibnamefont{and} \bibinfo{author}{\bibfnamefont{A.~C.}
  \bibnamefont{Gossard}}, \bibinfo{journal}{Nano Lett.}
  \textbf{\bibinfo{volume}{5}}, \bibinfo{pages}{1285} (\bibinfo{year}{2005}).

\bibitem[{\citenamefont{Metalidis and Bruno}(2005)}]{bruno}
\bibinfo{author}{\bibfnamefont{G.}~\bibnamefont{Metalidis}} \bibnamefont{and}
  \bibinfo{author}{\bibfnamefont{P.}~\bibnamefont{Bruno}},
  \bibinfo{journal}{Phys. Rev. B} \textbf{\bibinfo{volume}{72}},
  \bibinfo{pages}{235304} (\bibinfo{year}{2005}).

\bibitem[{\citenamefont{Thomas et~al.}(1996)\citenamefont{Thomas, Nicholls,
  Simmons, Pepper, Mace, and Ritchie}}]{pepper}
\bibinfo{author}{\bibfnamefont{K.~J.} \bibnamefont{Thomas}},
  \bibinfo{author}{\bibfnamefont{J.~T.} \bibnamefont{Nicholls}},
  \bibinfo{author}{\bibfnamefont{M.~Y.} \bibnamefont{Simmons}},
  \bibinfo{author}{\bibfnamefont{M.}~\bibnamefont{Pepper}},
  \bibinfo{author}{\bibfnamefont{D.~R.} \bibnamefont{Mace}}, \bibnamefont{and}
  \bibinfo{author}{\bibfnamefont{D.~A.} \bibnamefont{Ritchie}},
  \bibinfo{journal}{Phys. Rev. Lett.} \textbf{\bibinfo{volume}{77}},
  \bibinfo{pages}{135} (\bibinfo{year}{1996}).

\bibitem[{\citenamefont{Asada et~al.}(2006)\citenamefont{Asada, Freyn, and
  Pichard}}]{afp}
\bibinfo{author}{\bibfnamefont{Y.}~\bibnamefont{Asada}},
  \bibinfo{author}{\bibfnamefont{A.}~\bibnamefont{Freyn}}, \bibnamefont{and}
  \bibinfo{author}{\bibfnamefont{J.-L.} \bibnamefont{Pichard}},
  \bibinfo{journal}{Eur.\ Phys. J. B} \textbf{\bibinfo{volume}{53}},
  \bibinfo{pages}{109} (\bibinfo{year}{2006}).

\bibitem[{\citenamefont{Freyn and Pichard}(2007{\natexlab{a}})}]{fp1}
\bibinfo{author}{\bibfnamefont{A.}~\bibnamefont{Freyn}} \bibnamefont{and}
  \bibinfo{author}{\bibfnamefont{J.-L.} \bibnamefont{Pichard}},
  \bibinfo{journal}{Phys. Rev. Lett.} \textbf{\bibinfo{volume}{98}},
  \bibinfo{pages}{186401} (\bibinfo{year}{2007}{\natexlab{a}}).

\bibitem[{\citenamefont{Freyn and Pichard}(2007{\natexlab{b}})}]{fp2}
\bibinfo{author}{\bibfnamefont{A.}~\bibnamefont{Freyn}} \bibnamefont{and}
  \bibinfo{author}{\bibfnamefont{J.-L.} \bibnamefont{Pichard}},
  \bibinfo{journal}{Eur. Phys. J. B} \textbf{\bibinfo{volume}{58}},
  \bibinfo{pages}{279} (\bibinfo{year}{2007}{\natexlab{b}}).

\bibitem[{\citenamefont{Datta}(1997)}]{datta}
\bibinfo{author}{\bibfnamefont{S.}~\bibnamefont{Datta}},
  \emph{\bibinfo{title}{Electronic Transport in Mesoscopic Systems}}
  (\bibinfo{publisher}{Cambridge University Press}, \bibinfo{year}{1997}).

\end{thebibliography}

\end{document}